\documentclass[prd,twocolumn,nofootinbib]{revtex4-1}
\usepackage{lipsum} 
\usepackage{multirow,xspace}
\usepackage{amsmath,amsfonts,amssymb}
\usepackage{graphicx}
\usepackage{ulem,fancyvrb}
\usepackage{hyperref}

\usepackage{cleveref}
\crefname{section}{Section}{Sections}
\crefname{equation}{Eq.}{Eqs.}
\crefname{figure}{Fig.}{Figs.}
\crefname{table}{Table}{Tables}

\usepackage[usenames,dvipsnames,svgnames,table]{xcolor}
\usepackage{xfrac}

\let\oldsqrt\sqrt
\def\sqrt{\mathpalette\DHLhksqrt}
\def\DHLhksqrt#1#2{%
\setbox0=\hbox{$#1\oldsqrt{#2\,}$}\dimen0=\ht0
\advance\dimen0-0.2\ht0
\setbox2=\hbox{\vrule height\ht0 depth -\dimen0}%
{\box0\lower0.4pt\box2}}

\DeclareMathOperator{\sgn}{sgn}

\newcommand{\mo}{\ensuremath{m_0}\xspace}
\newcommand{\mhf}{\ensuremath{\widetilde{m}_{\sfrac{1}{2}}}\xspace}
\newcommand{\az}{\ensuremath{A_0}\xspace}
\newcommand{\tb}{\ensuremath{\tan\beta}\xspace}

\newcommand{\TeV}{\ensuremath{\,\mathrm{Te\kern -0.08em V}}\xspace}
\newcommand{\GeV}{\ensuremath{\,\mathrm{Ge\kern -0.08em V}}\xspace}
\newcommand{\MeV}{\ensuremath{\,\mathrm{Me\kern -0.08em V}}\xspace}

\newcommand{\tev}{\ensuremath{\mathrm{Te\kern -0.08em V}}\xspace}
\newcommand{\gev}{\ensuremath{\mathrm{Ge\kern -0.08em V}}\xspace}

\newcommand{\na}{\ensuremath{\widetilde{\chi}_1^0}\xspace}

\newcommand{\smu}{\ensuremath{\widetilde{\mu}}\xspace}

\newcommand{\numu}{\ensuremath{\widetilde{\nu}_\mu}\xspace}

\newcommand{\bsmumu}{\ensuremath{B^0_s\to \mu^+\mu^-}\xspace}

\newcommand{\br}[1]{\ensuremath{\mathcal{B}r\left(#1\right)}\xspace}

\newcommand{\NEUAffil}{Department of Physics, Northeastern University, Boston, MA 02115, USA}

\allowdisplaybreaks[1]

\begin{document}

\title{Supersymmetry after the Higgs }

\author{Pran~Nath} 
\email[Email: ]{nath@neu.edu}
\affiliation{\NEUAffil}

\begin{abstract}
A brief review is given of the implications of a 126 GeV Higgs boson for the discovery of supersymmetry.
Thus a 126 GeV Higgs boson is problematic within the Standard Model because of vacuum instability
pointing to new physics beyond the Standard Model. 
The problem of vacuum stability  is overcome in  the SUGRA GUT model but 
the 126 GeV Higgs mass implies that the average SUSY scale lies in the several TeV region. 
The largeness of the SUSY scale relieves the tension on SUGRA models since it helps suppress flavor changing neutral
currents and CP violating effects and also helps in extending the proton life time arising from baryon and lepton
number violating  dimension five operators. The geometry of radiative breaking of the electroweak symmetry 
and fine tuning in view of the large SUSY scale are analyzed.Consistency with the Brookhaven $g_{\mu}-2$ result is discussed. It is also shown that a large SUSY scale implied by the 126 GeV Higgs boson mass allows for light 
gauginos (gluino, charginos, neutralinos) and sleptons. These along with the lighter third generation squarks
 are the prime candidates for discovery at RUN II of the LHC. Implication of the 126 GeV Higgs
boson for the direct search for dark matter is discussed.   Also discussed are the sparticles mass
hierarchies and their relationship with the simplified models under   the Higgs boson mass constraint. 
  \end{abstract}


\maketitle
In 2012 the Large Hadron Collider (LHC)  made a landmark discovery of a new  boson. Thus the 
 CMS and ATLAS collaborations discovered a  boson with a mass of $\sim 126$ GeV~\cite{CMS:2012ufa, ATLAS:2012tfa, CMS:2012nga, ATLAS:2012oga}.  It is now confirmed
 that this newly discovered particle  is the long sought after Higgs boson ~\cite{Englert:1964et, Higgs:1964ia, Higgs:1964pj, Guralnik:1964eu} which plays  a central role in the breaking of the electroweak symmetry. While the observed particle is the last missing piece of the Standard  Model there are strong indications that at the same time its discovery
 portends discovery of a new realm of physics specifically supersymmetry.  Below we elaborate on 
 this theme in further detail. \\
 
 The outline of the rest of the paper is as follows:  In \cref{sec1} we discuss the status of the Higgs boson
 in the Standard Model, the issue of vacuum instability and the need for new physics beyond the Standard 
 Model. In \cref{sec2} we consider the implications of a 126 GeV Higgs boson within the framework of 
 supersymmetry and specifically in the framework of supergravity unified models. As is well known a 126 GeV
 Higgs boson within supersymmetry leads to a high SUSY scale $M_s$ with $M_s$ lying in the 
 TeV region. On the other hand the Brookhaven $g_{\mu}-2$ experiment \cite{Bennett:2006fi} shows a $3\sigma$ deviation
 from the Standard Model prediction~\cite{Hagiwara:2011af, Davier:2010nc}. An effect of this size requires that the average scale of sparticle masses entering 
 the loops in the supersymmetric electroweak correction to $g_\mu-2$ be low, i..e, $O(100)$ GeV.  
  Assuming the $3\sigma$ effect is robust we discuss in \cref{sec3} how to reconcile the high SUSY scale that is indicated by the 126 GeV Higgs boson mass with the low SUSY scale indicated by the Brookhaven experiment.
  In \cref{sec4} we discuss the implications of the Higgs boson mass and the geometry of radiative  
electroweak symmetry breaking (REWSB). In \cref{sec5} we discuss the issue of fine tuning in view of the 
large SUSY scale implied by the Higgs boson mass.   The sparticle landscape is an important indicator of the underlying fundamental theory  and in \cref{sec6}  we discuss the sparticle landscape after the Higgs boson discovery. 
The connection of this landscape to the so called simplified models is also discussed.  The implications of the
Higgs boson mass at $126$ on the search for dark matter in direct detection is discussed in  \cref{sec7}.
Future prospects are discussed in \cref{sec8}. 
 \section{Higgs boson and New Physics \label{sec1}}
\begin{figure}[t]
$$
\includegraphics[width=0.3\textwidth]{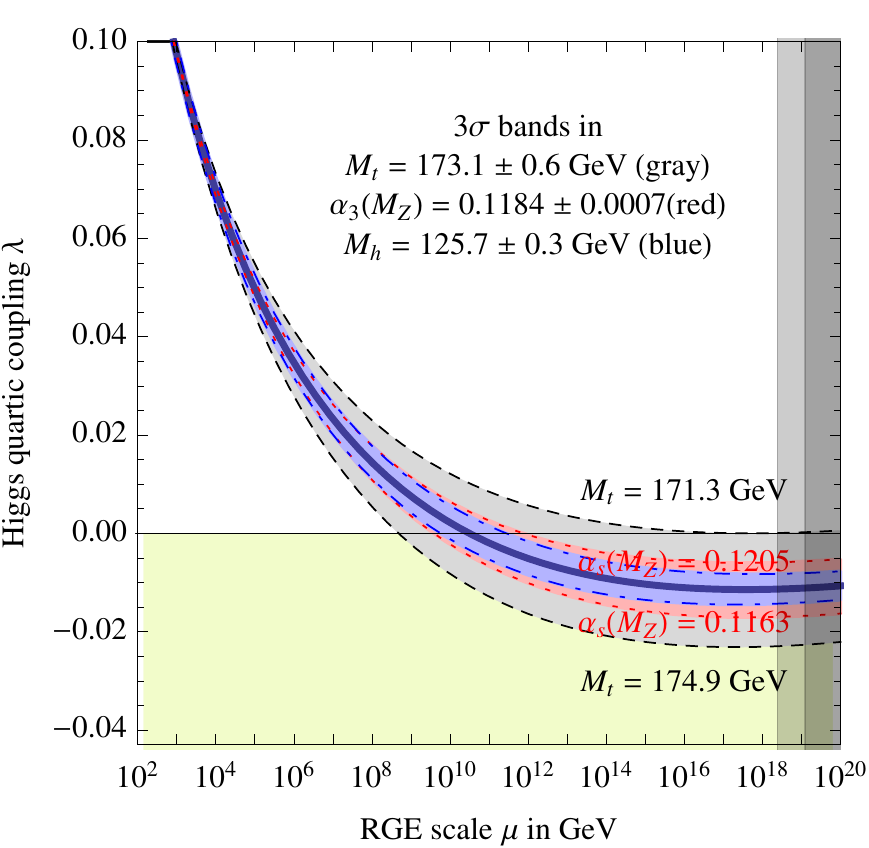}$$
\caption{
 RG evolution of the quartic coupling  $\lambda$ in the Standard Model 
varying 
$M_t$, $M_h$ and $\alpha_{\rm s}$ by $\pm 3\sigma$.
From \cite{Degrassi:2012ry}.}
\label{v-stability}
\end{figure}

  Within the Standard Model a Higgs boson mass of $\sim 126$ GeV is problematic.
  Thus renormalization group analyses show that vacuum stability up to the Planck scale  can be 
excluded at the 2$\sigma$ level for the Standard Model  for $M_h < 126$ GeV ~\cite{Degrassi:2012ry} as illustrated in \cref{v-stability}. Here one finds that the quartic Higgs boson coupling within 
the Standard Model turns negative at a scale around $10^{11}$ GeV making the vacuum unstable
~\cite{Degrassi:2012ry}.
However, as exhibited in \cref{v-stability} the result is rather sensitive to the mass of the top quark
(see also ~\cite{Masina:2012tz}). Thus lower values of the top mass tend to stabilize the vacuum while the higher values tend to destabilize it.
The current value of the top. i.e.,  $M_t= 173.21 \pm 0.51 \pm 0.71$ GeV suggests 
vacuum instability. Thus the discovery of $\sim 126$ GeV Higgs boson suggests the need for new physics 
beyond the Standard Model. Such new physics could be supersymmetry  and from here on we focus on this
possibility. 
\section{126 GeV Higgs  boson within SUSY \label{sec2}}
\begin{figure}[h!]
\centerline{
\includegraphics[width=6.cm]{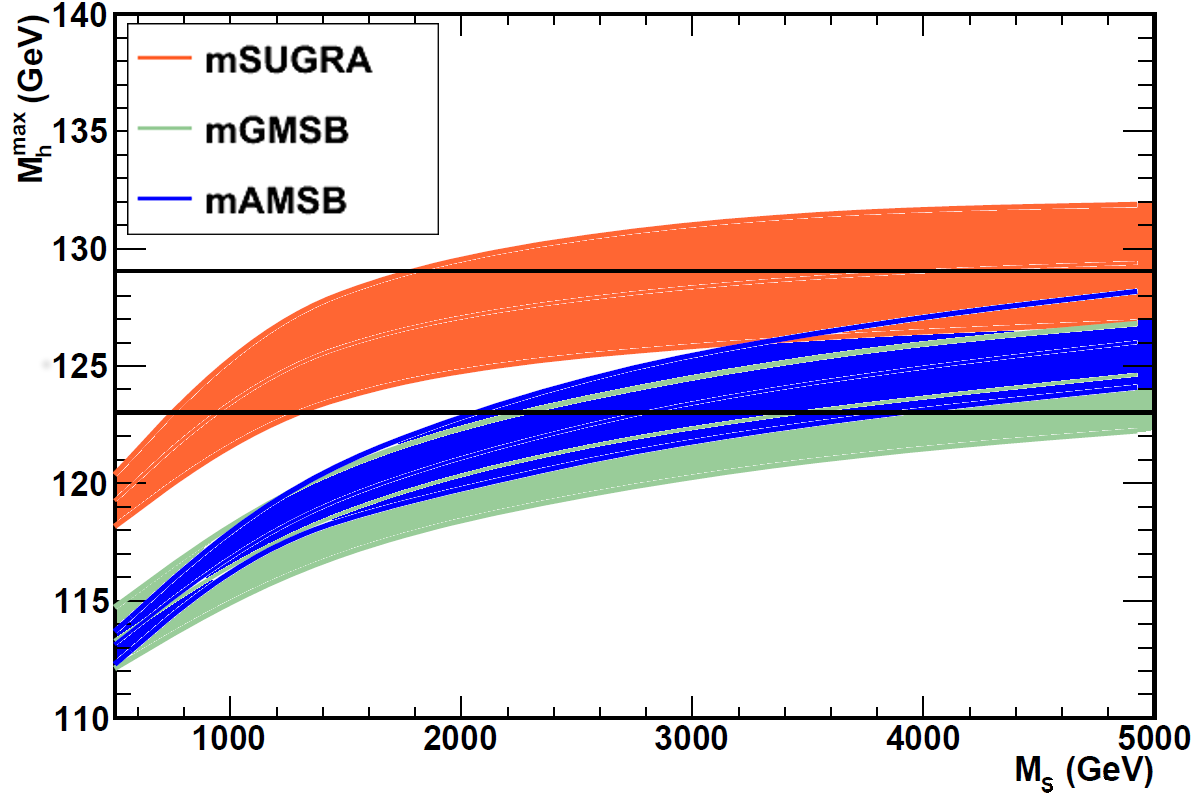}
}
\vspace*{0mm}
\caption
{The maximal Higgs mass in the constrained MSSM models 
mSUGRA, mGMSB and mAMSB (where mGMSM is the minimal gauge mediated symmetry breaking model
and mAMSB is the minimal anomaly mediated symmetry breaking model)
as a function of the SUSY scale $M_s$ when the top 
quark mass is varied in the range $M_t=\;$170--176 GeV.From ~\cite{Arbey:2012dq}.}
\label{djouadi-1}
\end{figure}

\begin{figure}[h!]
\hspace*{-.2mm}\includegraphics[width=6.5cm]{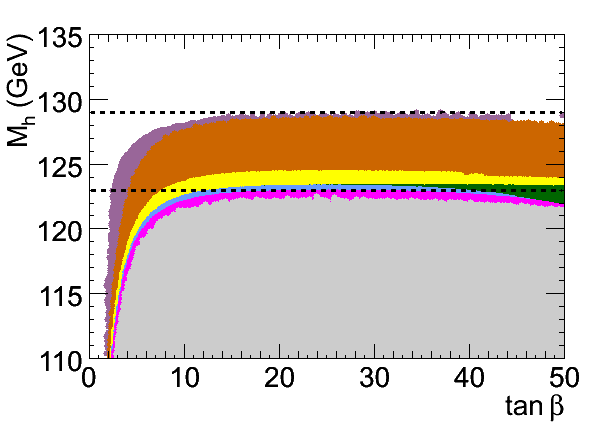}~~~\includegraphics[width=2.2cm]{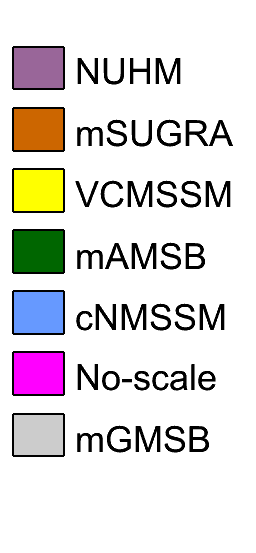}\\
\caption{The maximal Higgs mass  as a function of 
$\tan\beta$  in mGMSB,  mAMSB, and mSUGRA and  some of its variants.
The top quark mass is fixed at 
$M_t=173$ GeV. From ~\cite{Arbey:2012dq}.}
\label{djouadi-2}
\end{figure}

In contrast to the case of the Standard Model, in  supergravity unified models (SUGRA GUT) with the minimal supersymmetric standard model (MSSM) particle spectrum  vacuum stability can be  achieved up to high scales 
with color and charge conservation  (for a  recent analysis of vacuum stability after the Higgs boson discovery see \cite{Chowdhury:2013dka}).
However,  in MSSM  the lightest CP even Higgs boson has  a mass which lies below $M_Z$ ~\cite{Nath:1983fp,Gunion:1989we,Djouadi:2005gj,Carena:2002es}  and  a loop correction is needed  to pull its mass above 
$M_Z$ ~\cite{Berger:1989hg,Ellis:1990nz,Haber:1990aw,Okada:1990vk,Espinosa:2000df}.
Now the experimentally observed Higgs mass of  $\sim 126$ GeV requires a  large loop correction and correspondingly
a high  SUSY scale lying in the several TeV region ~\cite{Baer:2011ab, Arbey:2011ab, Draper:2011aa, Carena:2011aa, Akula:2011aa, Akula:2012kk, Strege:2012bt}. In fact the observed Higgs 
 mass is close to the upper limit predicted in grand unified supergravity models~\cite{Chamseddine:1982jx,Nath:1983aw,Hall:1983iz,Arnowitt:1992aq}  which predict an upper limit of around 130\GeV~\cite{Akula:2011aa,Akula:2012kk,Kane:2011kj, Arbey:2012dq,Ellis:2012aa,Baer:2012mv, Buchmueller:2013psa}
  (For recent reviews of Higgs and supersymmetry see~\cite{Nath:2012nh,Nath:2013okv}).
  It is possible to reduce the SUSY scale by including extra matter or from D term contributions from an extra $U(1)$ sector
 (see, e.g., \cite{Feng:2013mea} and the references therein). However,  we will focus on the implications 
 of the 126 GeV Higgs boson mass within the MSSM framework. Here it is instructive to ask if the Higgs
 boson mass measurement can shed some light on
   merits of the various  supersymmetry breaking mechanisms. 
 A relative comparison is given in \cref{djouadi-1} and
 \cref{djouadi-2}. 
  Specifically, the  mSUGRA analysis of  \cref{djouadi-1} and
 \cref{djouadi-2} uses  universal  boundary conditions of supergravity models 
  which are parametrized by the following set: 

 \begin{equation}
 m_0, m_{1/2}, A_0, \tan\beta, {\rm sign}(\mu) \,.
 \end{equation}
 Here $m_0$ is the universal scalar mass, $m_{1/2}$ is the universal gaugino mass, $A_0$ is the universal trilinear 
 coupling all at the grand unification scale, $\tan\beta =<H_2>/<H_1>$  where $H_2$ gives mass to the up quarks 
 and $H_1$ gives mass to the down quarks and the leptons and $\mu$ is the Higgs mixing parameter that 
 appears in the superpotential in the form $\mu H_1H_2$.  
 The analysis of \cref{djouadi-1} and
 \cref{djouadi-2} shows that a loop correction of the amount needed to pull the tree level value up to what is
 experimentally observed is easily achieved in mSUGRA. Further, a detailed analyses indicates
 that a sizable  $A_0$ is helpful in generating a large loop correction. \\
 

 Since a large higgs boson mass implies a large SUSY scale  it helps suppression of FCNC processes
 such as $b\to s\gamma$ \cite{Bertolini:1990if, Barbieri:1993av} and $B_s\to \mu^+\mu^-$.
Specifically it  explains why no deviations  in 
$B_s\to \mu^+\mu^-$ from the Standard Model result has been seen. 
Thus the LHCb collaboration ~\cite{LHCb:2012nna} determines the 
branching ratio \(\br\bsmumu = (3.2^{+1.5}_{-1.2})\times10^{-9} \), which is in excellent agreement with 
the Standard Model, which implies that the supersymmetric contribution~\cite{Choudhury:1998ze, Babu:1999hn, Bobeth:2001sq,Arnowitt:2002cq}
to this decay be very small. We note that the supersymmetric contribution to this decay
is mediated by  the neutral Higgs bosons and involve a flavor-changing scalar quark loop.
It is also sensitive to $CP$ violation~\cite{Ibrahim:2002fx, Ibrahim:2007fb}. If the SUSY scale is large,
the flavor-changing squark loop is suppressed. Further, the supersymmetric contribution is proportional
to $\tan^6\beta$ and  large values of $\tan\beta$, i.e., as large as $\sim 50$ tend to significantly 
enhance the 
supersymmetric contribution. For the same reason more moderate values of $\tan\beta$  would rapidly
bring down the supersymmetric contribution.
Thus a lack of a small deviation of the $B_s\to \mu^+\mu^-$
branching ratio from the Standard Model result is easily understood in the context of a large SUSY scale
and a moderate value of $\tan\beta$.\\

  Further, a large SUSY scale helps stabilize the proton against
decays from the baryon and lepton number violating dimension five operators. This is so because proton
decay via  baryon and lepton number violating dimension five operators involves 
dressing loop diagrams with  sparticle exchanges. Very crudely the proton decay from these is proportional 
to $m_{\chi_1}^2/m_{\tilde q}^4$ where $\chi_1^{\pm}$ is the chargino and the $\tilde q$ is the squark.
 Thus a large sfermion mass will lead to a desirable suppression of 
proton decay and an enhancement of its lifetime. As indicated in the preceding discussion 
 there is a strong
correlation between the SUSY scale and the Higgs mass. This strong correlation also implies a strong 
correlation between the proton lifetime from dimension five operators and the Higgs boson mass.
This  correlation is illustrated in Fig. 4. The current data also puts a lower limit on the heavier Higgs boson
mass (see, e.g., ~\cite{Nath:2012nh,Maiani:2012qf}).
\begin{figure}[t!]
\begin{center}
\includegraphics[scale=0.20]{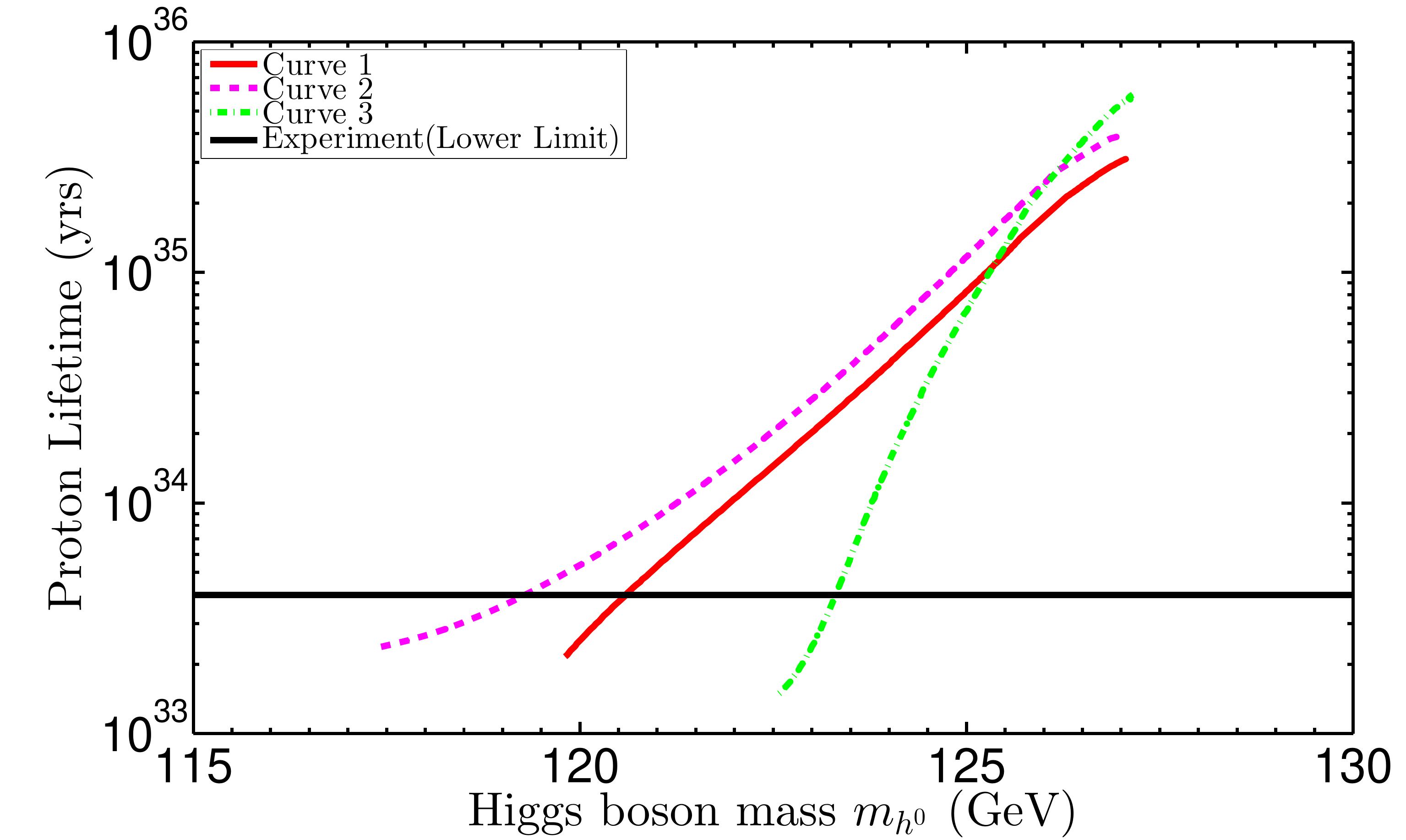}\\
\caption{
Dependence of the partial proton decay lifetime 
$\tau(p\to \bar \nu K^+)$ arising from baryon and lepton number violating dimension five operators 
 on $m_{h^0}$. 
Top  curve: $m_{1/2}=4207$ GeV, $A_0=20823$GeV, $\tan\beta=7.3$.
Middle:
 $m_{1/2}=2035$GeV, $A_0=16336$GeV, $\tan\beta=8$,
Bottom: $m_{1/2}=3048$GeV, $A_0 / m_0  =  -0.5$, $\tan\beta=6.5$. For all cases $M_{H_3}^{\rm{eff}}/M_G=50$
where $M_{H_3}^{\rm{eff}}$ is the Higgs triplet mass.  From \cite{Liu:2013ula}.   }
\end{center}
\label{fig:pdecay}
\end{figure}
\section{ Higgs boson mass and $g_\mu-2$  \label{sec3}}
While the Higgs boson mass of 126 GeV indicates a high SUSY scale in the TeV region, the 
Brookhaven experiment  on $g_{\mu}-2$ points to a rather low SUSY scale. Thus the Brookhaven Experiment E821~\cite{Bennett:2006fi} finds that 
\(a_\mu = \frac{1}{2}(g_\mu-2)\) deviates from the Standard Model prediction~\cite{Hagiwara:2011af, Davier:2010nc} at the $3\,\sigma$ level. Supposing that the entire deviation comes  from a supersymmetric contribution one 
finds 
\begin{equation}
  a_\mu^\mathrm{SUSY} = \delta a^{\rm exp}_\mu = (287 \pm 80)\times10^{-11} ~.
  \label{BNL}
\end{equation}
It has long been known that sizable contributions to $g_{\mu}-2$ can arise from the supersymmetric electroweak
sector~\cite{Yuan:1984ww, Kosower:1983yw,Lopez:1993vi,Chattopadhyay:1995ae,Moroi:1995yh,Ibrahim:1999aj,Heinemeyer:2003dq,Sirlin:2012mh}. These contributions arise from the  $\chi^{\pm}-$\numu and  from 
the $\chi^0-$\smu loops. While the detailed analysis is  involved,  a rough
approximation of the supersymmetric correction is given by 
\begin{equation}
  \delta a_{\mu} \simeq \sgn(M_2 \mu) \left(130 \times 10^{-11}\right) \left(\frac{100\GeV}{M_s}\right)^2 \tb ~,
  \label{g-2}
\end{equation}
where $M_s$  is the average SUSY scale that enters the loops.
From \cref{g-2} one finds that the average SUSY scale $M_s$ must be relatively small
to make a contribution of the size given by \cref{BNL}.\\

The above discussion points to an apparent tension between the SUSY scale indicated by the LHC measurement 
of the Higgs boson and the Brookhaven observation of a $3\sigma$ deviation.  The following possibilities
arise to reconcile the above two results. First it could be that the Higgs boson mass receives corrections
from extra matter (see, e.g., \cite{Feng:2013mea} and the references there in).
 There is another possibility which is that the Higgs boson could receive D term contributions 
from an extra $U(1)$ which is present in many extensions of the Standard Model.  In principle this would 
allow one to have a low SUSY scale. However, one must also fold in the fact that the LHC has not
observed any sparticles thus far. Specifically this puts the squarks and the gluino masses in the TeV range.\\

\begin{figure}[h!]
\begin{center}
\includegraphics[scale=0.2]{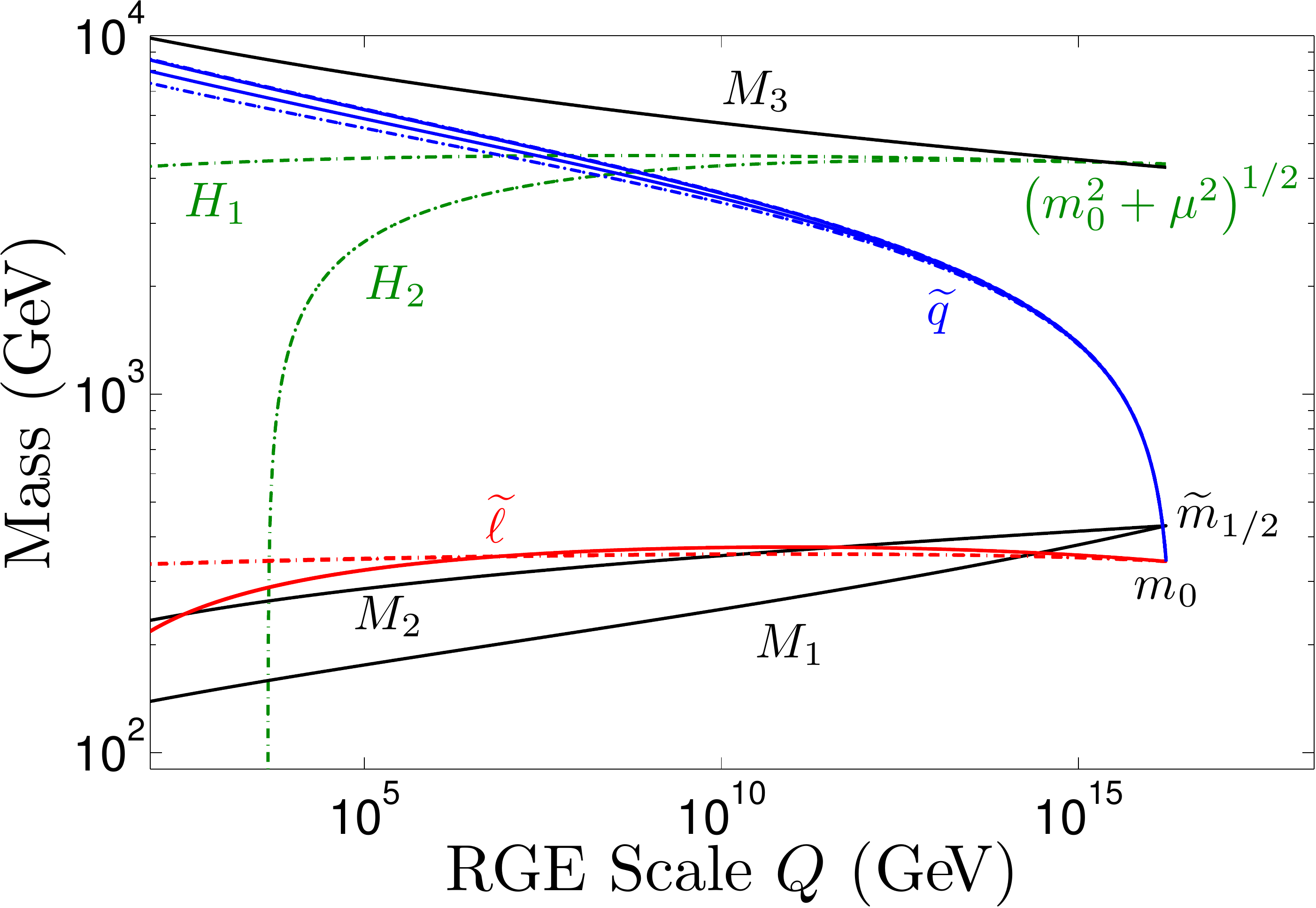}\\~\\
\end{center}
\caption{An illustration of the evolution of sparticle masses in the $\tilde g$SUGRA model. 
The analysis shows 
that starting with a  universal scalar mass at the GUT scale, the renormalization group
evolution splits the squark masses from the slepton masses by huge amounts.
From \cite{Akula:2013ioa}.
}
\label{RGE}
\end{figure}

\begin{figure}[h!]
\begin{center}
\includegraphics[scale=0.4]{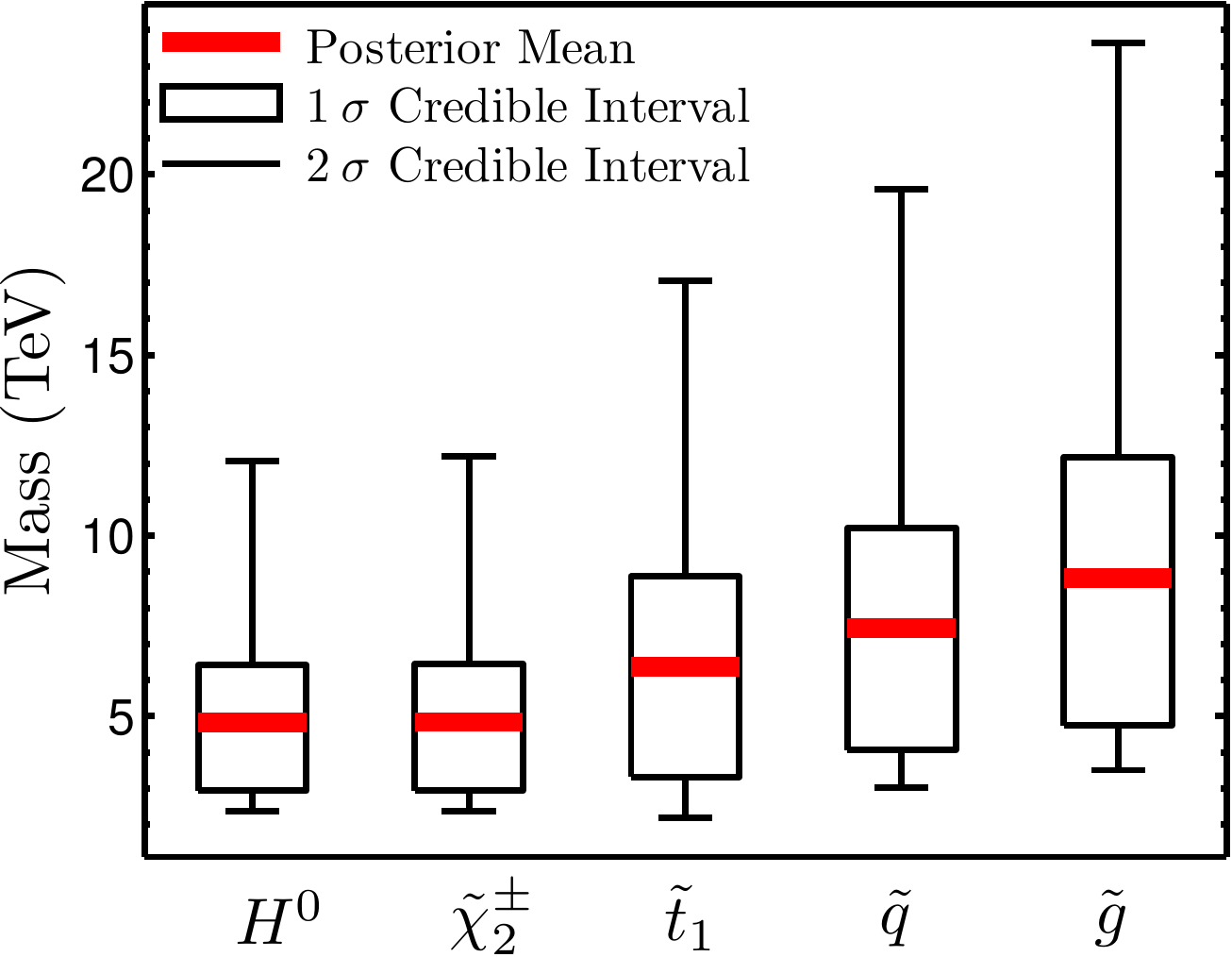}
\includegraphics[scale=0.4]{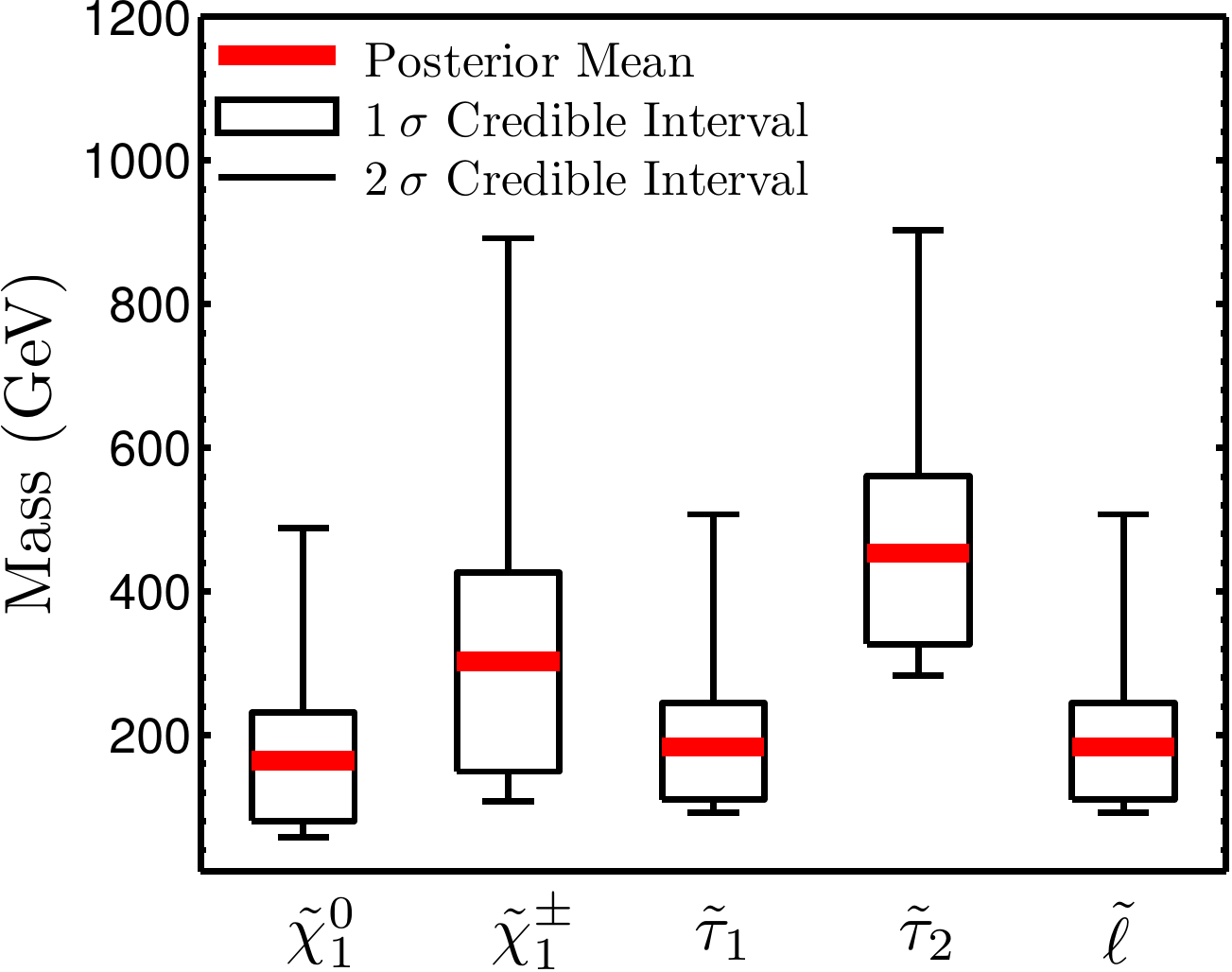}\\
\end{center}
\caption{Top panel: An exhibition of the heavy sparticle spectrum in the $\tilde g$SUGRA model.
Bottom panel: An exhibition of the light sparticle spectrum in the $\tilde g$SUGRA model. 
From \cite{Akula:2013ioa}.}
\label{split-scale}
\end{figure}

We discuss now a mechanism by which one can resolve the tension between the LHC result and the Brookhaven experiment.
This can be accomplished by a minimal extension of the supergravity model with universal boundary conditions. Thus 
suppose we consider  a supergravity model with universal scalar mass and universal trilinear coupling but with 
non-universality in the gaugino sector.  Specifically we consider the  boundary conditions on soft parameters to be
given by ~\cite{Akula:2013ioa} 
\begin{gather}
m_0, \tilde m_{1/2},  m_3, A_0, \tan\beta, {\rm sign}(\mu)\,,
\end{gather}
where $m_0<<m_3,  m_1=m_2=\tilde m_{1/2}<< m_3$. As an illustrative example we choose $m_3/m_1=10$. 
In this case the gluino mass enters in the RG evolution of the squarks and being 
the largest mass drives the RG evolution which results in radiative breaking of the electroweak symmetry
(For a review see \cite{ibanez:2007pf}). We label this model $\tilde g$SUGRA since REWSB is driven by the
gluino mass. 
 Specifically we exhibit the evolution of the masses of the third generation 
squarks $\tilde U$, $\tilde Q$ and the Higgs $H_2$ which form a coupled set due to the large Yukawa coupling
as shown in \cref{rgevolution}. 
\begin{gather}
\frac{d}{dt}\left[\begin{array}{ccc}
         m^2_{H_2}           \\
         m_{\tilde U}^2    \\
         m_{\tilde Q}^2        
	\end{array}\right] 
= - Y_t  \left[\begin{array}{ccc}
        3      &  3 & 3           \\
        2      &   2 &   2  \\
        1 &  1 & 1 \\
	\end{array}\right]
\left[\begin{array}{ccc}
        m^2_{H_2}           \\
         m_{\tilde U}^2    \\
         m_{\tilde Q}^2        
	\end{array}\right]
		 -Y_t A_t^2 \left[\begin{array}{ccc}
         3           \\
         2   \\
         1        
	\end{array}\right]	\nonumber\\
	+ 
		  \left[\begin{array}{ccc}
 3\tilde\alpha_2m_2^2+
\tilde\alpha_1m_1^2                   \\
{ \frac{16}{3}\tilde\alpha_3m_3^2}
+ \frac{16}{9} \tilde\alpha_1m_1^2           \\
{\frac{16}{3}\tilde\alpha_3m_3^2}
+ 3\tilde\alpha_2m_2^2+
\frac{1}{9} 
\tilde\alpha_1m_1^2                
	\end{array}\right].
	\label{rgevolution}
\end{gather}
Here  $Y_t=h_t^2/(4\pi^2)$ where $h_t$ is the top Yukawa coupling and $A_t$ is the trilinear 
coupling in the top sector. 
Suppose $m_3$ is in the TeV range but $m_0, m_1, m_2={\cal O} (100)$ GeV. In this case the RG 
evolution drives  squarks to become heavy with masses in the several TeV region
while the sleptons remain light. A numerical analysis of the RG evolution
is given in Fig. \ref{RGE}.  Here one finds that the evolution of  sfermion masses which start with a universal
scalar mass at the GUT scale generates  a significant split as the masses evolve towards the electroweak 
scale. Thus  at the electroweak scale the squark masses lie in the several TeV region due to the 
large contributions of the gluino mass term in the RG evolution while the slepton masses do not receive
large contributions from the gluino mass and remain relatively light. Further, since $m_1,m_2 << m_3$ the
electroweak gauginos all remain light. Thus one has a split scale SUSY which is not to be confused with the
split SUSY model~\cite{ArkaniHamed:2004fb}.
 The spectrum of the split scale SUSY is exhibited in Fig. \ref{split-scale}. The analysis Fig. \ref{split-scale}
 was done using a Bayesian statistical analysis (for other works using statistical approach see 
 \cite{deAustri:2006pe,Feroz:2008xx,Kim:2013uxa}). 
  From the lower panel of Fig. \ref{split-scale} we see that the electroweak gauginos $\tilde \chi_1^{0}$ and
$\tilde \chi_1^{\pm}$ as well as the sleptons are  all light. This light spectrum allows one to have sizable 
 supersymmetric electroweak contributions to $g_{\mu}-2$.  At the same time since the squarks are heavy
 as seen from the upper panel of Fig.\ref{split-scale} the SUSY scale that enters in the loop correction to
 the Higgs boson is substantial which produces the desired size correction to the Higgs boson mass. 
For related works see~\cite{Giudice:2012pf,Ibe:2013oha,Mohanty:2013soa,Bhattacharyya:2013xba,Babu:2014lwa}. 

\section{ Higgs Boson Mass and  REWSB Geometry \label{sec4}}
The Higgs boson at $\sim 126$ GeV and its implication that the SUSY scale is high has important implications 
for the geometry of radiative breaking of the electroweak symmetry.
It was pointed out long time ago that REWSB  in general contains two main branches, the ellipsoidal branch (EB)
and the hyperbolic branch (HB)~\cite{Chan:1997bi,Chattopadhyay:2003xi,Baer:2003wx}
(For related works see \cite{Feng:1999mn,Cassel:2009ps,Feldman:2011ud}).
 To understand the relationship 
it is convenient to write the REWSB constraint that determines $\mu^2$ in the following form
\begin{align}
\label{mus}
 \mu^2 & = 
\left( \begin{matrix}
+1	~~~~({\rm EB})\\
  ~0~~~~~~({\rm FP}) 	\\
  -1 ~~~~({\rm HB}) 
\end{matrix} \right)
  ~ {m^2_0}  ~|C_1|+  \Delta (m_{1/2}, A_O, \tan\beta) \,,\\  
    +1&:  {\rm Euclidean ~geometry} \Rightarrow {\rm EB ~of ~REWSB}\,, 
     \label{plusone}\\
  -1&:    {\rm Hyperbolic ~geometry} \Rightarrow {\rm HB ~of ~REWSB} \,,
\label{minusone}  
\end{align}
where $\Delta$ is typically  positive and   where $C_1$ is a function that depends only on $\tan\beta, M_t$ and the renormalization group scale $Q$ 
but does not depend on $m_0, m_{1/2}, A_0$.  
The signs $(\pm)$ in \cref{mus}
are determined by the sign of $C_1(Q)$ with depends on the renormalization group scale $Q$.
The RG scale $Q$ is chosen so that the two loop correction to the scalar potential is minimized. Typically
$Q\sim M_s$ where $M_s= \sqrt{m_{\tilde t_1} m_{\tilde t_2}}$. \\~\\

\begin{figure}[h!]
\begin{center}
\includegraphics[scale=0.15]{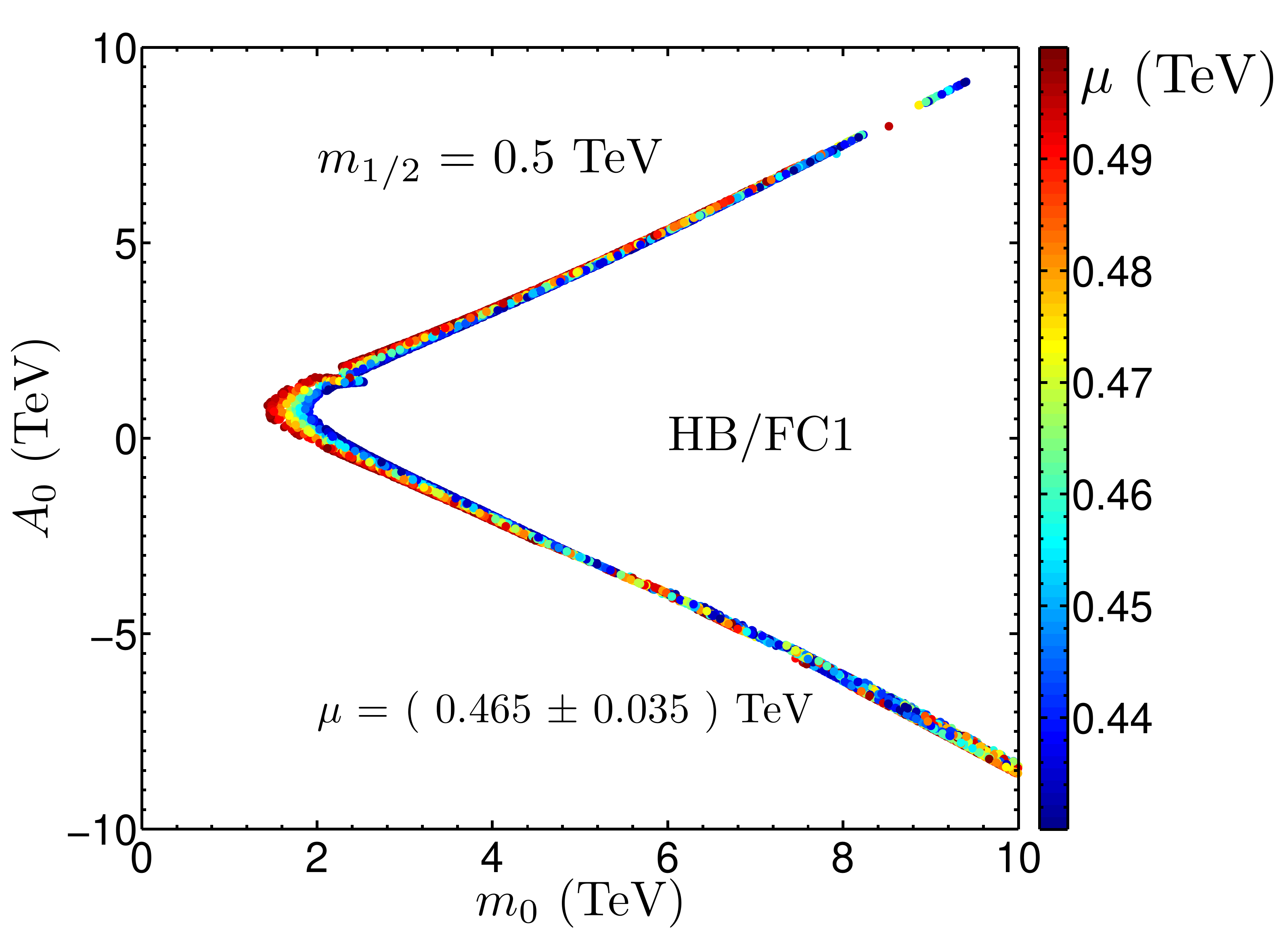}\\
\includegraphics[scale=0.2]{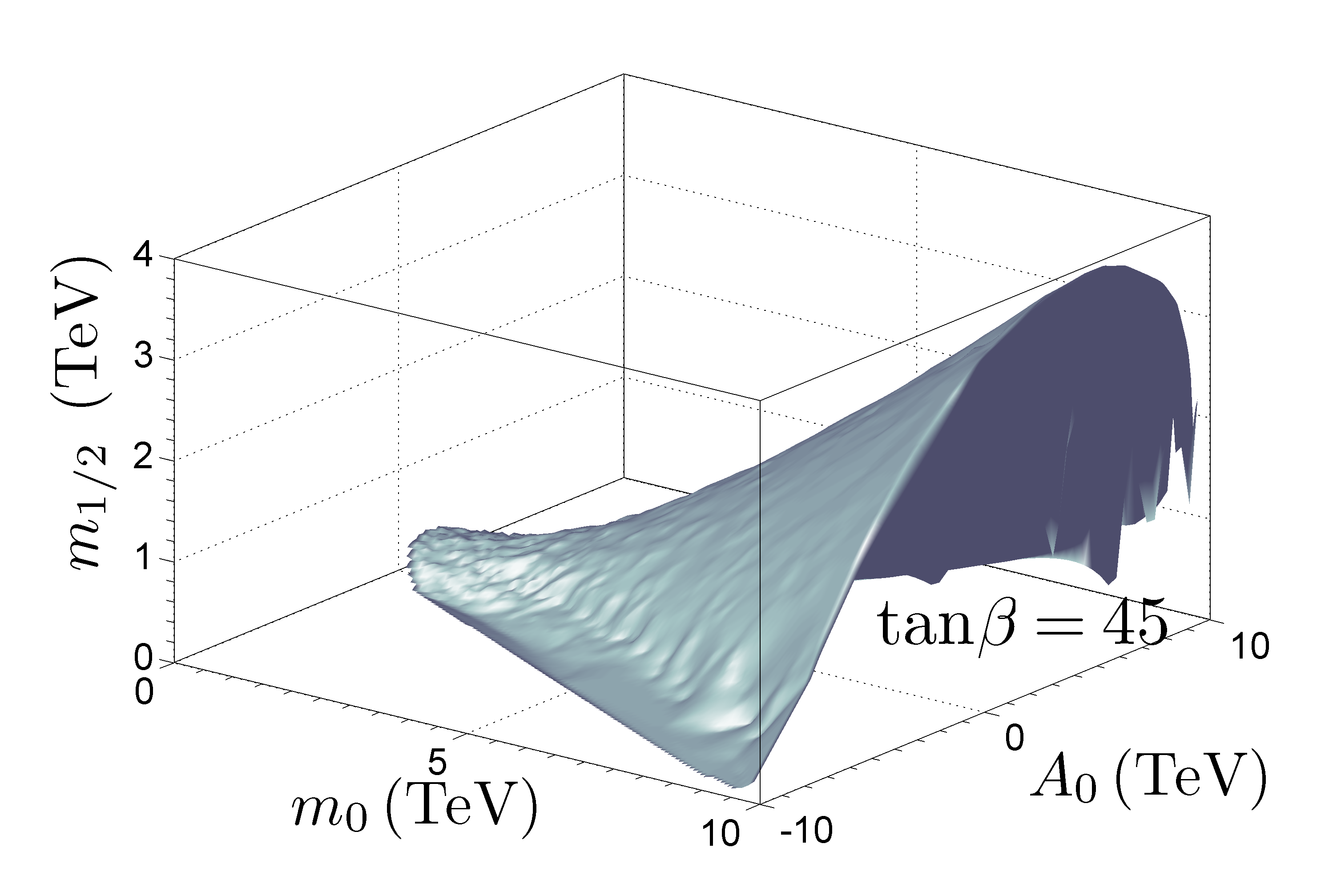}
\end{center}
\caption{Top panel: An illustration of the Focal Curve region of the hyperbolic branch. 
Bottom panel: An illustration of the Focal Surface region of the hyperbolic branch.
 From~\cite{Akula:2011jx}. }
\label{focalregions}
\end{figure}

Let us discuss now the implications of \cref{mus}. First consider the case when $C_1=0$. In this case 
one finds that at the one loop level,  $\mu^2$ no longer depends on $m_0$. Thus here $m_0$ can get
large without affecting $\mu$. This is the focus point region (FP)~\cite{Feng:1999mn}. However, a small $\mu$ can be gotten 
in other ways even when $C_1$ is non-vanishing. Thus suppose $C_1$ is non-vanishing and negative.
In this case one can have cancellation between the $C_1$ term and the $\Delta$ term so that 
$\mu^2$ is still small. In fact the curves of fixed $\mu^2$ in the plane of two of the other parameters
are hyperbolic curves  or focal curves. An illustration is given  in the upper panel of  \cref{focalregions}.
An illustration of the focal surface region of the hyperbolic branch when one considers the  3-d region of $m_0, m_{1/2}, A_0$ 
is given in the upper panel of  \cref{focalregions}. 
 Finally suppose $C_1$ is positive. In this 
case as $m_0$ gets large $\mu$ also gets large since $\Delta$ is typically positive. Thus the region of small $\mu$ is not
achieved in this case. This is the ellipsoidal branch (EB). In summary on EB a large $m_0$ implies a large $\mu$.
On FP and on HB  a small $\mu$ can be consistent with a 
large $m_0$. Phenomenologically a small $\mu$ is desirable since it allows one to have all the EW gauginos 
 in the sub TeV range. A small $\mu$ is also desirable for observation of dark matter. \\~\\
 
        \section{Fine tuning \label{sec5}}
    The  Higgs boson mass of 126 GeV and the large SUSY scale it implies raise the issue of naturalness and
    fine tuning which we address below (for a related work see~\cite{Baer:2012cf}).   It should be kept in mind, however,
    that the issue of what constitutes  fine tuning is a rather subjective one and depends in part  on 
   what phenomena are included in the analysis. Generally, fine tuning analyses include
just the constraints of radiative breaking of the electroweak symmetry.
 However, it is not unreasonable to include other phenomena such as 
suppression of flavor changing neutral currents and  CP violating effects~\cite{Jaeckel:2012aq}
and  suppression of  proton decay from dimension five operators~\cite{Arnowitt:1993pd,Liu:2013ula} within the
framework of grand unified theories. 
As an illustration of how the fine tuning is affected by including more than just REWSB constraints, 
we consider a fine tuning analysis including two different phenomena, i.e., REWSB and proton stability.
We  will consider fine tuning  for these two phenomena  separately and then combine them to produce
a composite fine tuning. \\

Thus to define fine tuning for the REWSB one considers the minimization of the  scalar potential
  that determines the Z -boson mass and one has 
\begin{equation}
\frac{1}{2}M_Z^2= -\mu^ 2 + |m_{H_u}|^2 + \cdots
\end{equation}
Here one notices that as  $\mu$ or $|m_{H_u}|$ gets large,  one needs a fine tuning to get the  $Z$ mass down
to the experimental value. 
 An  obvious way to define fine tuning  then  is via the ratio $F$ given by  
 \begin{equation}
 F\sim \frac{ 2 |m_{H_u}|^2}{M_Z^2} \,.
 \label{rewsb1}
 \end{equation}
 \cref{rewsb1} indicates that low values of $m_{H_u}$ and  thus  low values of soft masses 
 $O(M_Z)$ will lead to low values of fine tuning. \\

However, low values of soft masses can create problems as they lead to large  FCNC and CP violating 
effects. Further, low values of soft masses will in general lead to rapid 
proton decay from baryon and lepton number violating dimension five operators in unified models of particle interactions
(for recent reviews see~\cite{Nath:2006ut,Raby:2008pd,Babu:2013jba}).
 The fine tunings using FCNC and CP violating effects has been discussed specifically in ~\cite{Jaeckel:2012aq}
 and here we focus on proton decay. 
 Thus to get the proton lifetime prediction from baryon and lepton number violating dimension five operators
 to be consistent with the experimental upper limit on $p\to \bar \nu K^+$  partial lifetime
  one needs a fine tuning of input
 parameters and in this case  one may define fine tuning so that 
\begin{gather}
F_{pd} = \frac{4\times 10^{33} {\rm yr}}{\tau(p\to \bar \nu K^+){\rm yr}}\,.
\label{5.4}
\end{gather}
As indicated in \cref{sec2}
very roughly $\tau(p\to \bar \nu K^+)\sim C(m_{\chi^\pm}/m_{\tilde q}^2 M_{H_3}^{eff})^{-2}$
where $M_{H_3}^{eff}$ is the Higgs triplet mass. 
\cref{5.4} makes  clear that larger squark masses  will lead to a 
larger proton lifetime and a smaller value of fine tuning parameter $F_{pd}$. In general one may have 
several such fine tuning parameters $F_i$, where $i$ takes on the values $1\cdots n$. 
In this case a more appropriate object to consider is a composite fine tuning $\cal F$ defined by ~\cite{Liu:2013ula}
\begin{equation}
{\cal F}=\left(\prod_{i=i}^n F_i\right)^{\frac{1}{n}} \,.
\label{5.5}
\end{equation}
An analysis of fine tunings defined by \cref{rewsb1}, \cref{5.4}  and \cref{5.5} 
is given in Fig. 8 of ~\cite{Liu:2013ula}.
Here the red region indicates fine tuning when
one considers just the radiative electroweak symmetry breaking. As expected the fine tuning increases with 
increasing $m_0$.  The blue region gives the fine tuning for the proton decay arising from dimension five
operators. As expected the fine-tuning here is a falling function of $m_0$. The composite fine tuning defined via
\cref{5.5} is given by the black region. One finds that the overall fine tuning decreases with a large $m_0$
and thus a smaller fine tuning  occurs at large $m_0$. This result is in contrast to the conventional view that a small
SUSY scale $O(M_Z)$  constitutes  a small fine tuning.  The analysis above shows that inclusion of flavor 
and proton stability constraints argue for an overall  larger SUSY scale. 
\begin{table}[h!]
\begin{center}
\begin{tabular}{|l|l|c|}
\hline
Pattern Label
		&  \multicolumn{1}{c|}{Mass Hierarchy}
			& \%	 \\
\hline
\hline
mSP[C1a]
		& $\chi^{\pm}_{1} < \chi^{0}_{2} < \chi^{0}_{3}< \chi^{0}_{4}$
			&  83.8 \\
mSP[C1b]
		& $\chi^{\pm}_{1} < \chi^{0}_{2} < \chi^{0}_{3}< H^0$
			&  2.49 \\
mSP[$\tau$1a]
		& $\tau_{1} < \chi^{0}_{2} <\chi^{\pm}_{1}< H^0 $
			&  3.89 \\
mSP[N1a]
		& $ \chi^{0}_{2} < \chi^{\pm}_{1} < H^0 < A^0$
			&  3.31 \\
\hline
\end{tabular}
\caption{\label{tab2}
A sample of sparticle mass hierarchies for mSUGRA, where $\chi_1^0$ is the LSP. The input at the GUT scale consists of 
  $\mo \in [0.1,10] \TeV$, $m_{1/2} \in [0.1,1.5] \TeV$, $\frac{\az}{\mo} \in [-5,5]$, $\tan\beta \in [2,50]$, $\mu > 0$, with the constraints $\Omega h^2 < 0.12$, {$m_{h^0} > 120 \GeV$}.    Here and in Table II and  Table III the last column gives the percentage with which the patterns appear
in the scans.  From  \cite{Francescone:2014pza}.}
\end{center}
\label{tab1}
\end{table}
\section{Higgs and the sparticle landscape \label{sec6}}
It has been demonstrated in several previous works that 
sparticle mass hierarchies can be used as discriminants of the underlying models of SUSY 
breaking~\cite{Feldman:2008hs,Feldman:2007fq,Chen:2010kq,Feldman:2007zn,Conley:2010du,Altunkaynak:2010tn}.
However, 
 the landscape of mass hierarchies is large.
    Thus there are 31 additional particles
 beyond the spectrum of the Standard Model and there are a priori $31!$ ways in which these particles
 can hierarchically  
 arrange themselves. Using Sterling's formulae, i.e..  $n!\sim \sqrt{2 \pi n} (n/e)^n$, 
 one finds that $n=31$  gives $\sim 8\times 10^{33}$ possibilities. 
 Actually the possible landscape of sparticle 
mass hierarchies  is even larger since 
 the mass gaps among the sparticle masses can vary continuously which makes
the allowed sparticle landscape even larger than the string landscape of  $\sim 10^{500}$
string vacua. Truncated hierarchies (e.g., n=3,4) are much smaller and one can generate 
a list of simplified models   from these.  In previous works an analysis of the sparticle mass hierarchies was
carried out where the experimental lower limits of the Higgs mass constraint given by LEP II was used. 
The observation of the Higgs boson mass at $\sim 126$ GeV imposes a  much more severe constraint 
and one might ask what the set of allowed mass hierarchies is in this case. A full analysis of this issue
is given in \cite{Francescone:2014pza}
 where the implications for the mSUGRA case as well as for the 
SUGRA models with non-universalities  (nuSUGRA)  are investigated [The literature on non-universalities in 
supergravity models is extensive. For a  sample see 
~\cite{Ellis:1985jn,Drees:1985bx,Nath:1997qm,Ellis:2002wv,Anderson:1999uia,Anderson:1996bg,Huitu:1999vx,Corsetti:2000yq,
Chattopadhyay:2001mj,Chattopadhyay:2001va,Martin:2009ad,Feldman:2009zc,Gogoladze:2012yf,Ajaib:2013zha,Kaufman:2013oaa}  and for a review see~\cite{Nath:2010zj}].
 Here we give a brief description of the main results.\\
 
The analysis of mass hierarchies under the Higgs boson mass constraint for the mSUGRA model is given
in  Table 1 where we display only those mass hierarchies which have a frequency of occurrences larger than
2\%.  The sequence $\chi^{\pm}_{1} < \chi^{0}_{2} < \chi^{0}_{3}< \chi^{0}_{4}$ implies that the mass of $\chi_1^0$
is smaller than the mass of $\chi_2^0$, the mass of $\chi_2^0$ is smaller than the mass of $\chi_3^0$ and so on.
Here the constrain on the relic density of $\Omega h^2 < 0.12$,  and on the Higgs boson mass of $m_{h^0} > 120 \GeV$
were imposed. A similar analysis for the non-universal supergravity models with  non-universality in the 
$SU(2)_L$ gaugino masses is displayed in Table 2, while the analysis with non-universality in the gluino masses
is displayed in Table III.\\

\begin{figure}[t]
  \begin{center}
{\rotatebox{0}{\resizebox*{8cm}{!}{\includegraphics{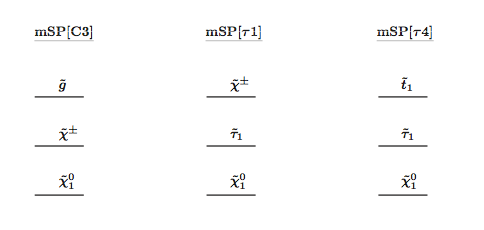}}\hglue5mm}}\\
   \end{center}
   \caption{An illustration of three simplified models that arise from truncation of  UV complete models.
   From  \cite{Francescone:2014pza}.   
   }
   \label{simplified}
\end{figure}

\begin{table}[t!]
\begin{center}
{\footnotesize
\begin{tabular}{|l|l|c|}
\hline
Pattern Label
		&  \multicolumn{1}{c|}{Mass Hierarchy}
			& \%	 \\
\hline
\hline
nuSP$_2$[C1a]
		& $ \chi^{\pm}_1 < \chi^{0}_2 < \chi^{0}_3 < \chi^{0}_4 $
			&  35.71 \\
nuSP$_2$[C1b]
		& $ \chi^{\pm}_1 < \chi^{0}_2 < \chi^{0}_3 < \chi^{\pm}_2 $
			&  14.57 \\
nuSP$_2$[C2a]
		& $ \chi^{\pm}_1 < \chi^{0}_2 < \tau_1 < \nu_\tau $
			& 12.25  \\ 
nuSP$_2$[C3a]
		& $ \chi^{\pm}_1 < \chi^{0}_2 < g < \chi^{0}_3 $
			&  6.237 \\ 			
nuSP$_2$[C3b]
		& $ \chi^{\pm}_1 < \chi^{0}_2 < g < t_1 $
			&  4.820 \\ 
nuSP$_2$[C4a]
		& $ \chi^{\pm}_1 < \chi^{0}_2 < H^0 < A^0$
			&  2.767 \\ 
nuSP$_2$[C5b]
		& $ \chi^{\pm}_1 < \chi^{0}_2 < t_1 < \tau_1$
			&  2.746 \\ 
nuSP$_2$[N1a]
		& $ \chi^0_2 < \chi^\pm_1 < g < \chi^0_3$
			&  	2.246	 \\
\hline
\multicolumn{3}{c}{\vspace{-1 mm}}  \\
\end{tabular}
}
\caption{\label{nusugra_lightchargino_mass_patterns}\label{tab4}
A sample of sparticle mass hierarchies for the non-universal SUGRA
(nuSUGRA) model with a light chargino. 
The high scale parameters lie in the range  $\mo \in [0.1,10] \TeV$, $M_1=M_3=\mhf \in [0.1,1.5] \TeV$, $M_2 = \alpha \mhf$, $\alpha \in [\frac{1}{2},1]$, $\frac{\az}{\mo} \in [-5,5]$, $\tan\beta \in [2,50]$, $\mu > 0$, with the constraints $\Omega h^2 < 0.12$, {$m_{h^0} > 120 \GeV$}.   From  \cite{Francescone:2014pza}}
\end{center}
\label{tab2}
\end{table}
\begin{table}[t!]
\begin{center}
{\footnotesize
\begin{tabular}{|l|l|c|}
\hline
Pattern Label
		&  \multicolumn{1}{c|}{Mass Hierarchy}
			& \%	 \\
\hline
\hline
nuSP$_3$[C1a]
		& $ \chi^\pm_1 < \chi^0_2 < \chi^0_3 < \chi^0_4 $
			& 	63.273	 \\
nuSP$_3$[C1b]
		& $ \chi^\pm_1 < \chi^0_2 < \chi^0_3 <  g $
			& 	10.263	 \\
nuSP$_3$[C1c]
		& $ \chi^\pm_1 < \chi^0_2 < \chi^0_3 <  H^0$
			& 	4.587	 \\
nuSP$_3$[C1d]
		& $ \chi^\pm_1 < \chi^0_2 < \chi^0_3 < \tau_1 $
			& 	4.243	 \\
nuSP$_3$[C1e]
		& $ \chi^\pm_1 < \chi^0_2 < \chi^0_3 < t_1 $
			& 	4.549	 \\
nuSP$_3$[g1a]
		& $ g < \chi^0_2 < \chi^\pm_1 < \chi^0_3 $
			& 	3.940	 \\	
nuSP$_3$[g1b]
		& $ g < \chi^0_2 < \chi^\pm_1 < t_1 $
			& 	3.031	 \\	
\hline
\multicolumn{3}{c}{\vspace{2.65cm}}  \\
\end{tabular}
}
\vspace{-2.5cm}
\caption{A sample of sparticle mass hierarchies for the nuSUGRA model with a light gluino. 
The high scale parameters  lie in the range 
 $\mo \in [0.1,10] \TeV$, $M_1=M_2=\mhf \in [0.1,1.5] \TeV$, $M_3 = \alpha \mhf$, $\alpha \in [\frac{1}{6},1]$, $\frac{\az}{\mo} \in [-5,5]$, $\tan\beta \in [2,50]$, $\mu > 0$, with the constraints $\Omega h^2 < 0.12$, {$m_{h^0} > 120 \GeV$}.
 From  \cite{Francescone:2014pza}. 
}
\end{center}
\label{tab3}
\end{table}
One may note that  the so called simplified models~\cite{ArkaniHamed:2007fw,Alwall:2008ag,Alwall:2008va,Alves:2011sq,Alves:2011wf,Barnard:2014joa,Kraml:2013mwa,Edelhauser:2014ena}
can be generated from the analysis of UV complete models presented in Tables I-III 
and those discussed in ~\cite{Francescone:2014pza}  by a truncation. Thus  a truncation of SUGRA GUTs keeping  a few  lowest mass particles generates  a large number of simplified models.
An illustration  is given in \cref{simplified} where three simplified models are shown that
arise from UV complete models under the Higgs boson mass constraint and under the relic density constraint. 
Currently many experimental analyses are being done using simplified models. However, it should be 
kept in mind that these models are only truncations of the more complete models which should be 
used for comparison with the underlying theory that generates these mass hierarchies.

\vspace{0cm}
\section{Higgs and dark matter\label{sec7}}
\begin{figure}[h!]
\begin{center}
{\rotatebox{0}{\resizebox*{8.5cm}{!}{\includegraphics{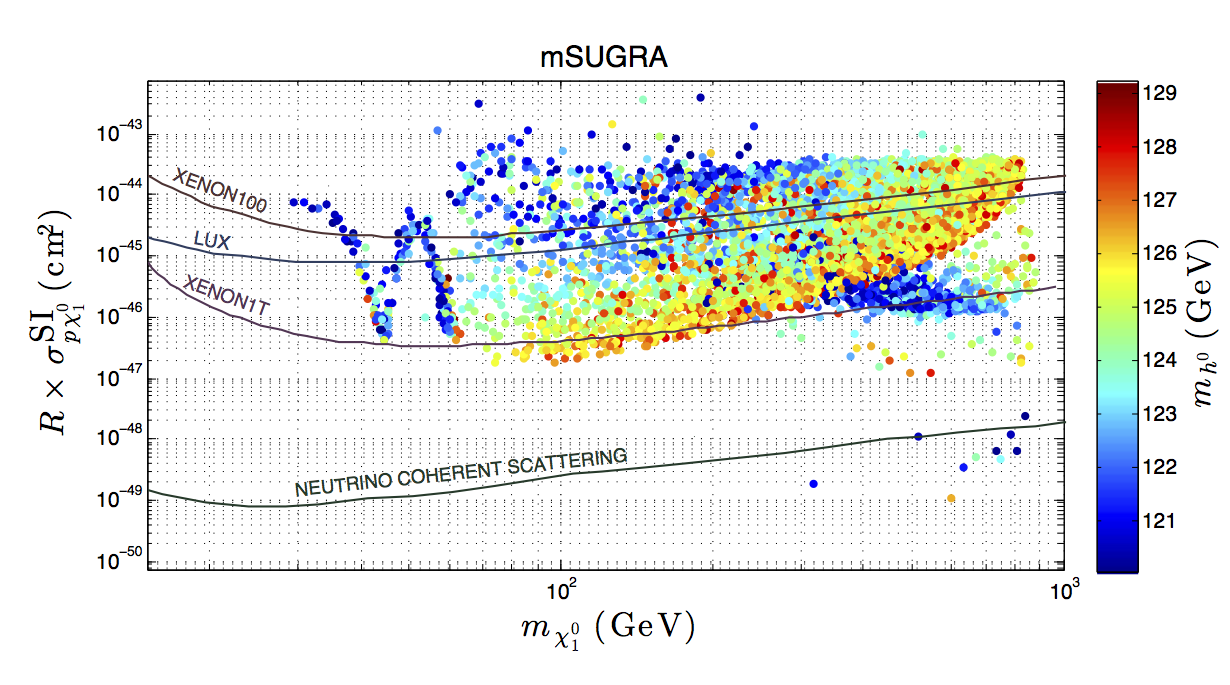}}\hglue5mm}}
\end{center}
\caption{An exhibition of the spin independent neutralino-proton cross section $R\times \sigma^\text{SI}_{p,\chi_1^0}$ vs the neutralino mass for mSUGRA where the colors exhibit the Higgs boson mass. From  \cite{Francescone:2014pza}.
}
\label{darkmatter}
\end{figure}

The Higgs boson mass constraint also has  significant implications for 
dark matter detection (for a related work see \cite{Roszkowski:2014wqa}).
An analysis of the spin independent neutralino-proton cross section 
$R\times \sigma^\text{SI}_{p,\na}$, where $R= (\Omega h^2)_{\rm theory}/(\Omega h^2)_{\rm WMAP}$,
as a function of the
neutralino mass is given in \cref{darkmatter} where the parameter points are colored according to the 
Higgs boson mass in the  mass range 120 GeV -129 GeV.   Here the deep blue corresponds to the smallest and
the deep red to the largest Higgs boson mass. One may notice that most of the parameter points 
with the Higgs mass in the vicinity of the values measured by the ATLAS and CMS experiments lie below 
the current lower limit of the LUX experiment~\cite{Ghag:2014uva}. The analysis further shows that future dark matter experiments
such as XENON1T \cite{Aprile:2012zx} will be able to explore a large part of the parameter space consistent with the measured
Higgs boson mass.

\section{Future prospects \label{sec8}} 
In view of the TeV size  SUSY scale indicated by the Higgs boson mass, LHC RUNII, which will likely operate
at $\sqrt s=13$ TeV,  has a better chance of observing sparticles than LHC7+8, at least those sparticles  which are low lying.
These low lying sparticles are mostly uncolored particles, and in addition the gluino and the lightest  squarks
are also possible candidates for discover. Recent analyses have shown that LHC RUN II could observe
 gluinos up to $\sim 2$ TeV ~\cite{Baer:2009dn,Cohen:2013xda}  and the CP odd Higgs up to around a TeV 
~\cite{Altunkaynak:2013xya}.
Further hints of new physics beyond the Standard Model can come from precision measurement of
the Higgs boson couplings to fermions at the International Linear Collider, a high energy $e^+e^-$ machine 
which can measure most of the Higgs-fermion couplings to an accuracy of up to 5\% \cite{Peskin:2012we}.
For the direct observation of heavier sparticles one  will require  a hadron collider with much larger energies
than those at LHC RUNII,  such as a
 100 TeV hadron collider. Analyses indicate that such a machine could detect a 
 gluino up to as much as $\sim 10$ TeV  as indicated by simulations based on  simplified  models ~\cite{Cohen:2013xda}.\\

  \begin{figure}[h!]
\begin{center}
{\rotatebox{0}{\resizebox*{6cm}{!}{\includegraphics{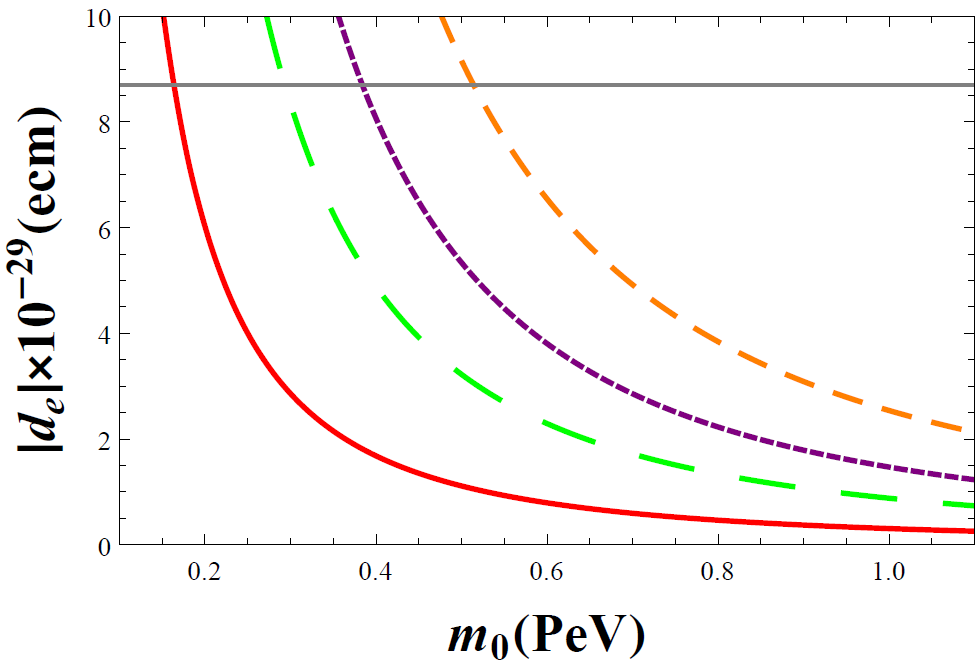}}\hglue5mm}}\\
\caption{An illustration of the probe of high SUSY scales using the electron EDM.  
The plot exhibits  the electron EDM as a function of $m_0$  for different values of the phase $\alpha_\mu$ of the 
Higgs mixing parameter $\mu$.  The curves are for 
 the cases $\alpha_\mu= -3$ (small-dashed, red), $\alpha_\mu=-0.5$ (solid), $\alpha_\mu=1$ (medium-dashed, orange), and $\alpha_\mu= 2.5$ (long-dashed, green).  The other parameters are  $\text{$|\mu |$ = }4.1\times 10^2\text{ ,  $|$}M_1\text{$|$ = }2.8\times 10^2\text{ ,  $|$}M_2\text{$|$ = }3.4\times 10^2\text{ ,  $|$}A_e\text{$|$ = }3\times 10^6\text{ ,  }m_0^{\tilde{\nu}}\text{ = }4\times 10^6\text{ ,  $|$}A_0^{\tilde{\nu}}\text{$|$ = }5\times 10^6\text{ , tan$\beta $ = }30$ . All masses are in GeV,  phases in rad and EDM in $e$cm.The analysis shows that improvements in the electron EDM constraint can probe scalar masses in the 100 TeV- 1 PeV region and beyond.
 The top horizontal line is the current experimental limit from the ACME Collaboration~\cite{Baron:2013eja}.
  From \cite{Ibrahim:2014tba}.}
\label{edm}
\end{center}
\end{figure}

Further precision experiments can allow one to probe even higher 
SUSY mass scales.~\cite{McKeen:2013dma,Moroi:2013sfa,Altmannshofer:2013lfa,Ibrahim:2014tba,Dhuria:2014fba}.
One example is to use EDMs as a probe of high SUSY scales.  Thus the electron EDM is
 most stringently constrained  by the ACME Collaboration ~\cite{Baron:2013eja} which gives
\begin{equation}
 |d_e| < 8.7\times  10^{-29} ~e{\rm cm}\ .
\end{equation}
 \cref{edm} exhibits the dependence of the electron EDM on $m_0$ for a number of CP phases of the 
 Higgs mixing parameter.
The electron EDM limit is  likely to improve by an order of magnitude in the coming years
and thus the future EDM measurements will allow one to extend the probe of new physics  
 up to a PeV or  more  as shown in \cref{edm}. 
 One can also 
use the precision measurement of $g-2$ of the electron as a sensitive probe of new physics~\cite{Giudice:2012ms,Aboubrahim:2014hya}.\\

In summary the discovery of the Higgs boson and the measurement of its mass at $\sim 126$ GeV has 
very significant implications for new physics beyond the Standard Model. The fact that a  126 GeV Higgs boson 
mass  in the framework of the Standard Model makes the vacuum unstable provides yet another 
reason why new physics beyond the Standard Model must exist.
 The most promising candidate for such
physics is supersymmetry. Specifically within the concrete framework of supergravity grand unification
one finds that the Higgs boson mass is predicted to lie below $\sim 130$ GeV. 
The fact that the observed HIggs boson mass respects this bound is a significant support for 
SUGRA GUT. Further, the 
Higgs boson mass of $\sim 126$ GeV requires the average SUSY scale to be high, i.e., in the TeV region.
This high scale explains why we have seen no significant deviation from the Standard Model prediction
 in FCNC processes such as $b\to s\gamma$ and $B_s \to \mu^+ \mu^-$. Further, the same high SUSY scale
 explains the non-observation of sparticles in $\sqrt s=7$ TeV and $\sqrt s= 8$ TeV data at RUN I of the LHC. \\

 As discussed in \cref{sec3}
 one important issue pertains to the Brookhaven experiment which sees a $3\sigma$ deviation in 
 $g_{\mu}-2$ from the Standard Model  prediction. This effect is difficult to understand within the supergravity
 model with universal boundary conditions. However, it is not difficult to explain the observed phenomenon
 within supergravity unified models with non-universal boundary conditions.  Here it is possible to 
 have light electroweak gauginos and light sleptons while the squarks are heavy.  
  In this case one can explain the Brookhaven $g_\mu-2$ result as well as achieve a Higgs
 boson mass consistent with experiment. The discovery of the Higgs boson mass is important not only because one
 has found the  last missing piece of the Standard Model  but also because it is likely the first piece of a new class of models such as supersymmetric models  which require the existence of a whole new set of particles. It is hoped that LHC RUN II will reveal some of these. 
 
\vspace{1.0cm}
\begin{acknowledgments}
This research is  supported in part by grants  NSF grants PHY-1314774 and PHY-0969739,  
and by XSEDE grant TG-PHY110015. This research used resources of the National Energy Research Scientific Computing 
Center, which is supported by the Office of Science of the U.S. Department of Energy under Contract No. DE-AC02-05CH11231 
\end{acknowledgments}

\bibliography{master}

\end{document}